\documentclass{nature}

\usepackage{epsfig,amsmath}
\usepackage{graphicx}
\usepackage{color}
\usepackage[10pt]{moresize}
\usepackage{bm}
\usepackage{SIunits}

\let\oldcaption\caption
\renewcommand{\caption}{\sffamily \oldcaption}
\begin{document}

\title{Experimental Determination of the Finite-Temperature Phase Diagram of a Spin-Orbit Coupled Bose Gas}

\author{Si-Cong Ji$^{1,3,\dagger}$, Jin-Yi Zhang$^{1, 3,\dagger}$, Long Zhang$^{1,3}$, Zhi-Dong Du$^{1,3}$, Wei Zheng$^2$,
You-Jin Deng$^{1,3}$, Hui Zhai$^2$, Shuai Chen$^{1,3,\star}$ and Jian-Wei Pan$^{1,3,\star}$}

\maketitle

\begin{affiliations}
\item Shanghai Branch, Hefei National Laboratory for Physical Sciences at Microscale 
and Department of Modern Physics, University of Science and Technology of China,
Shanghai 201315, China
\item Institute for Advanced Study, Tsinghua University, Beijing, 100084, China
\item Synergetic Innovation Center of Quantum Information and Quantum Physics, 
University of Science and Technology of China, Hefei, Anhui 230026, China
\end{affiliations}

\clearpage

\begin{abstract}

Spin-orbit (SO) coupling has led to numerously exciting phenomena in electron systems. Whereas the synthesized SO coupling with ultracold neutral atoms gives us an opportunity to study SO coupling in bosonic systems, which exhibit many new phenomena of superfluidity and various symmetry breaking condensate phases. A richer structure of symmetry breaking always results in a nontrivial finite-temperature phase diagram, however, the thermodynamics of the SO coupled Bose gas at finite temperature is still unknown so far either in theory or experiment. Here we experimentally determine a novel finite temperature phase transition that is consistent with a transition between the stripe ordered phase and the magnetized phase. We also observe that the magnetic phase transition and the Bose condensate transition occur simultaneously as temperature decreases. Our work determines the entire finite-temperature phase diagram of SO coupled Bose gas and demonstrates the power of quantum simulation.

\end{abstract}



Superfluidity is a phenomenon known for century in physics but the study of superfluidity still
keeps producing novel physics. Recently SO coupling, which has played an important role in recently discovered topological insulator \cite{topological_insulator, topological_insulator2}, has also been realized in ultracold degenerate gases \cite{Lin2011, JZhang2012, MZwielein2012,jinyi2012,Spielman2012,chuanwei2013,long2013}. The SO coupled Bose gases are predicted to exhibit
a host of new phenomena of superfluidity. For instance, SO coupling leads to degenerate single-particle ground states,
which can result in a new type of stripe superfluid with spatial density order \cite{huizhai2010,Ho2011,yunli2012,Congjun}.
SO coupling can significantly enhance low-energy density-of-state that dramatically increases quantum and thermal
fluctuation effects and also magnifies interaction effects \cite{Jian,Lamacraft,Baym,Cui,Zheng,Kamenev}.
The absence of Galilean invariance due to SO coupling yields unconventional behavior of superfluid
critical velocity \cite{Biao,Strigari2012,Zheng}.

In this work we generate SO coupling in $^{87}$Rb Bose gases by two contour-propagating laser beams
as described in previous works \cite{Lin2011,jinyi2012}. In this setup only the motion along the spatial direction
of Raman laser (denoted by $\hat{x}$) is coupled to spin, and the single-particle Hamiltonian along $\hat{x}$ is given by ($\hbar=1$)
\begin{equation}
\hat{H}_0=\frac{\left(k_x-k_{\rm r}\sigma_z\right)^2}{2m}+\frac{\delta}{2}\sigma_z+\frac{\Omega}{2}\sigma_x.
\end{equation}
We focus on the case with $\delta=0$ 
where the system has an additional $Z_2$ symmetry ($k_x\rightarrow-k_x$ and $\sigma_z\rightarrow -\sigma_z$ simultaneously).
The single-particle dispersion is shown in Fig. 1(a). 
To motivate our study of finite-temperature physics, we shall first summarize what are known at zero temperature.

For $\Omega<\Omega_2\simeq 4E_{\rm r}$ ($E_{\rm r}=k_{\rm r}^{2}/(2m)$), there are two degenerate single-particle
minima denoted by $\pm k_{\text{min}}$ and their wave functions are represented by $\psi_{\text{L}}$ and $\psi_{\text{R}}$,
respectively, and these two degenerate states have opposite magnetization.
Due to this degeneracy, wave function of Bose condensation should be determined by interactions in this regime.
Theoretical results \cite{Ho2011,yunli2012} have shown that for interaction parameters of ${}^{87}$Rb atoms,
the condensate wave function is in a superposition state $(\psi_{\text{L}}+\psi_{\text{R}})/\sqrt{2}$
for $\Omega<\Omega_1\simeq 0.2E_{\rm r}$ and bosons condensate either into $\psi_{\text{L}}$ or into $\psi_{\text{R}}$ for $\Omega_1<\Omega<\Omega_2$.
For the former, condensate displays periodic density stripe order and the spatial translational symmetry is spontaneously broken.
For the latter, spatial transitional symmetry is not broken but the $Z_2$ symmetry is broken,
and condensate is therefore magnetic.
Experimentally, although the stripe order has not been directly imaged so far,
miscible to immiscible transition has been observed at $\Omega \simeq \Omega_1$ \cite{Lin2011}.
Finally, when $\Omega>\Omega_2$, the single-particle dispersion only has one single minimum at zero momentum,
and its wave function denoted by $\psi_0$ displays zero magnetization.
A divergent spin susceptibility has been observed at $\Omega=\Omega_2$ \cite{jinyi2012}.

In short, as shown in Fig. 1(a), at zero temperature, the system will undergo two successive magnetic phase transitions
as $\Omega$ increases, first from non-magnetic stripe (ST) phase to magnetized (MG) phase,
 and then from MG  phase to non-magnetic zero-momentum (NM) condensate.
We remark that the ST phase at $\Omega<\Omega_1$ and the MG phase at $\Omega>\Omega_1$ are two fundamentally different phases,
since they exhibit very different behaviors in terms of magnetization, symmetry breaking and low-energy excitation spectrums.
At zero temperature, phase boundary between them is determined by interaction energy only \cite{Ho2011,yunli2012}.
While finite-temperature phase diagram should contain richer physics due to the interplay between interaction effects
and thermal effects, it has not been explored either theoretically or
experimentally so far.
One question is which phase will be more favorable as temperature increases.
Three different sceneries are presented in Fig 1(b), namely, ST phase first turns into MG phase
before becoming normal phase (b1), or ST phase directly turns into normal phase (b2),
or MG turns into ST before becoming normal phase (b3).
Another fundamental question is that in the regime where the low-temperature phase is a MG phase,
whether the spontaneous magnetization ($Z_2$ symmetry broken) happens simultaneously with spontaneous phase coherence
(Bose condensation) transition. If these two transitions occur in two different temperatures,
exotic state can be expected in the temperature window between these two transitions \cite{Z2,Lamacraft1,Lamacraft2}.
In this paper we experimentally explore these questions and also present our physical explanation.

First, we would like to determine the critical temperature $T_{\text{c}}$ and address the transition between normal Bose gas and condensate.
A two-dimensional bi-mode fitting function with dispersion of SO coupled bosons is applied to fit the TOF image (For detail, see Supplementary Material).
The temperature of the SO coupled Bose gas in the experiment is determined by fitting the wing of TOF image,
as shown in the lower part of Fig 2(a).
The total number of thermal atom $N_{\text{th}}$ of the two spin states is also determined from the fitting.
Meanwhile, the total number of atoms $N$ is obtained directly from counting atoms in the image.
Thus, the number of condensed atom is given by $N_{\text{0}}=N-N_{\text{th}}$.
The condensate fraction $f=N_{\text{0}}/N$ is obtained.
For a given Raman coupling $\Omega$, we plot condensate fraction $f$ as a function of temperature $T$ for a range of small $f$,
as shown in Fig 2(b). An empirical formula $f(T)=1-(T/T_{\text{c}})^{\alpha}$ for $T<T_{\text{c}}$
and $f(T)=0$ for $T>T_{\text{c}}$ is applied to numerically fit the data.
Both $\alpha$ and $T_{\text{c}}$ are fitting parameters. The error bar of  $T_{\text{c}}$ is $3-5\%$, coming from
the uncertainty of both our measurement and fitting (For details of the error evaluation, see Supplementary Material).
Also at $T_{\text{c}}$, the total atom number $N$ and the trapping frequency $\omega$ are recorded.

In Fig 2(c) we plot the ratio of $T_{\text{c}}$ to the critical temperature $T_{\text{c}}^{({\rm s})}$ of single-component
ideal gas with same experimental parameters, which is immune to the variation of the atom number $N$ and trapping frequency $\bar\omega$ in the experiment (see Supplementary Material). 
At small $\Omega$, the system is close to a two-component thermal gas
and the density of each component is half compared to single-component one, thus, if one ignores interaction effects,
the ratio $T_{\rm{c}}/T_{\text{c}}^{({\rm s})}$ will drop to about $(1/2)^{1/3}\approx0.79$.
While when $\Omega>\Omega_2$, the dispersion has only one single minimum, 
the spin of all atoms will be polarized along $\hat{x}$  and the system will essentially become single component with effective mass\cite{spielmanPRA2009} $m^{*}=m\Omega/(\Omega-\Omega_2)$. 
Thus at large $\Omega$, $|T_{\rm{c}}|/T_{\text{c}}^{({\rm s})}$ increases toward unity. The red and blue curves in Fig. 2(c) represent the theoretical calculation of $T_{\text{c}}/T_{\text{c}}^{(\text s)}$ for atom number $N=1.0\times10^5$ and $1.0\times10^6$, respectively, with interaction effects taken into account.

Furthermore, we can study the interaction shift of $T_\text{c}$.
Experimentally, for each measurement we can determine $T_\text{c}$ and total number of atoms $N$.
With atom number $N$, trap parameter $\omega$ and SO coupling parameters $k_{\rm r}$ and $\Omega$,
the non-interacting critical temperature with trap $T^{(0)}_{\text{c}}$ can be calculated and then $\Delta T_{\text{c}}=T_\text{c}-T^{(0)}_{\text{c}}$ can be deduced. The experimental results of $\Delta T_\text{c}/T_\text{c}$ are shown in Fig 2(d). $\Delta T_\text{c}$ is always negative since repulsive interaction decreases density. When $\Omega<\Omega_2$, we find that $|\Delta T_\text{c}|/T_\text{c}$ increases from about $0.10$ at $\Omega\simeq 0$
to about $0.14$ at $\Omega\simeq\Omega_2$. And clearly when $\Omega>\Omega_2$, we find $|\Delta T_\text{c}|/T_\text{c}$ decreases
as $\Omega$ increases. 
The error bar of the data is transferred from the $T_{\text{c}}$ and the calculation of $T^{(0)}_{\text{c}}$.
For comparison, we plot two theory curves with different total atom numbers and trap frequencies,
because in experiment the total atom number and the trap frequency vary as $\Omega$.
These theory curves include both interaction effect with Hatree-Fock approximation
and trap effect with local-density approximation, and the detail of the calculation is given in Ref. \cite{Zheng}, as well as in Supplementary Material.
We find reasonable agreements between theory and experiment. This result shows that the interaction shift of
transition temperature reaches a maximum around $\Omega_2$, where the single-particle dispersion changes from double minimum
to a single-minimum and the low-energy density-of-state is maximized.

Here, we turn to study the magnetic property of the SO coupled Bose gas. The magnetization of the gas is defined as 
\begin{equation}
M_{i}=\frac{N_{i, \uparrow}-N_{i, \downarrow}}{N_{i, \uparrow}+N_{i, \downarrow}},
\end{equation}
Where $N_{i, \uparrow}$ and $N_{i, \downarrow}$ present atom number in $|\uparrow\rangle$ ( $|m_{F}=-1\rangle$)
and $|\downarrow\rangle$ ($|m_{F}=0\rangle$) states, respectively. The index $i=0$ ($i=\text{th}$) represents the condensate atoms (the thermal atoms). 
Experimentally, we repeat our measurements for thousands of times and record the probability of occurrence of different magnetization of the atoms to obtain the histograms for statistics. 

Now we focus on the regime with Raman coupling strength $\Omega<\Omega_1$ where the ground state at zero temperature is predicated as a ST condensate.
However, so far there is no direct experimental evidence of phase coherence between $\psi_{\text{L}}$ and $\psi_{\text{R}}$ or
direct image of stripe order in the ST phase. In this work, we shall distinguish ST phase and MG phase by
their magnetic properties. Let us first consider a uniform system without domain walls.
When the single-particle spectrum has two minima for $\Omega<\Omega_2$, $\psi_{\text{L}}$ and $\psi_{\text{R}}$
have opposite magnetization $M_{0}=\pm\sqrt{16-\Omega^{2}/E^2_{\rm r}}/4$.
Therefore, for MG condensate, the many-body wave function for condensate is either $\frac{1}{\sqrt{N_0!}}(\hat{a}^\dag_{\text{L}})^{N_0}|0\rangle$
or $\frac{1}{\sqrt{N_0!}}(\hat{a}^\dag_{\text{R}})^{N_0}|0\rangle$, where $N_0$ denotes total number of atoms in the condensate.
Thus, in the histogram of condensate magnetization, one expects two peaks at $M_{0}=\pm\sqrt{16-\Omega^{2}/E^2_{\rm r}}/4$. 
While in the presence of magnetic domains, that both spin up and spin down appear simultaneously while phase separated, the peaks will be smeared and the distribution could be flat. 
For the ST phase, the many-body wave function for condensate is given by
\begin{equation}
\frac{1}{\sqrt{N_0!}}\left(\frac{\hat{a}^\dag_{\text{L}}+\hat{a}^\dag_{\text{R}}}{\sqrt{2}}\right)^{N_0}|0\rangle=\frac{1}{\sqrt{N_0!}}\sum\limits_{m}c_m(a^\dag_{\text{L}})^{\frac{N_0}{2}-m}(a^\dag_{\text{R}})^{\frac{N_0}{2}+m}|0\rangle.
\end{equation}
with $c_m=\frac{(\frac{N_0}{2}-m)!(\frac{N_0}{2}+m)!}{N_0!}$.
Each measurement projects the coherent state into a Fock state $(a^\dag_{\text{L}})^{\frac{N_0}{2}-m}(a^\dag_{\text{R}})^{\frac{N_0}{2}+m}|0\rangle$ with a probability $\frac{c^2_m}{N_0!}$, with a magnetization $M_{0}=\frac{2m}{4N_0}\sqrt{16-\frac{\Omega^{2}}{E^2_{\rm r}}}$.
Since $c^2_m$ is a smooth function peaked at $m=0$, the histogram of condensate magnetization should also
show a distribution centered at $M_{0}=0$. Across a transition between ST and MG phase, one expects that the statistics of $M_{0}$ will behaves differently at the two sides. 

In this regime we choose three different Raman coupling strengths $\Omega=0.10E_{\rm r}$ (raw a), $0.15E_{\rm r}$ (raw b),
and $0.18E_{\rm r}$ (raw c), and the histograms of condensate magnetization are shown in Fig 3 for various temperatures
below $T_\text{c}$. Just below $T_{\text{c}}$, when the condensate fraction is very small ($0<f<0.02$),
it is very clear from Fig.~3 (a1,b1) and (c1) that the histogram shows two sharp peaks around $M\approx \pm 1$.
This evidence for the spontaneous $Z_2$-symmetry breaking of magnetization strongly supports
that the condensate is not in the ST phase but in the MG phase when the condensate first appears.
As further lowering temperature, we find that the peaks at $M\approx \pm 1$ gradually decreases,
and a board peak centered at $M=0$ gradually develops.
Finally at low temperature when the condensate fraction $f>0.25$, as shown in Fig.~3 (a6, b6) and (c6),
peaks at $M\approx \pm 1$ completely disappear and a Gaussian-like peak develops around $M=0$,
which is consistent with ST state in low temperature as analyzed above.
Comparing three different raws, we find that the closer to $\Omega_1$ is the Raman coupling, the lower is the transition temperature from MG to ST.
Roughly speaking, the transition from MG phase to ST phase takes place
when condensate fraction $f \approx 0.05$ for $\Omega=0.10E_{\rm r}$, $f \approx 0.10$ for $\Omega=0.15E_{\rm r}$,
and $f \approx 0.20$ for $\Omega=0.18E_{\rm r}$, as indicated by dashed lines in Fig.~3.
This supports that the finite-temperature phase diagram should be of b1 type among three scenarios presented in Fig 1(b),
i.e. as temperature increases, ST phase will first turn into a MG phase before becoming normal state.
This is reminiscent of transition from superfluid B-phase to superfluid A-phase as temperature increases
in Helium-3 \cite{He-3}. And a more similar analogy is the magnetic transition in undoped iron pnictide \cite{mag,mag1,mag2},
in which the low temperature phase is a spin-density-wave that breaks translational symmetry,
while as temperature increases, it undergoes a transition to spin nematic state
which restores transitional symmetry but breaks a discrete symmetry,
and then turns into paramagnetic state at a higher temperature.

To more quantitatively determine the phase boundary, we plot $\sqrt{\langle M_0^2\rangle}$
as a function of $f$ shown in the Fig.~3(d).
These plots all display a kink which we use to determine the location of phase boundary.
Thus our measurements put three points in the finite-temperature phase boundary between ST and MG phases, 
corresponding to the condensate fraction (measured temperature) of $f=0.06\pm0.02$ ($T=126\pm12 {\text{nK}}$) 
at $\Omega=0.10 E_{\text{r}}$, $f=0.09\pm0.02$ ($T=116\pm10{\text{nK}}$) at $\Omega=0.15E_{\text{r}}$, and $f=0.22\pm0.09$ ($T=62\pm27{\text{nK}}$) at $\Omega=0.18E_{\text{r}}$. 

Furthermore, we also carry out the magnetization histogram analysis at the lowest temperature of $T<20$ nK with a condensate fraction $f>0.9$. Zero-temperature theoretical calculations has determined the ST-MG transition at $\Omega \approx \Omega_1= 0.2E_{\text{r}}$ \cite{Lin2011,yunli2012} and experimental evidence of this transition has also been obtained in the lowest temperature \cite{Lin2011}.
As shown in Fig.4, the magnetization distribution for $\Omega<\Omega_1$ shows a single peak around $M_{0}=0$,
while the central peak gradually becomes flat and two peaks at $M_0 \approx \pm 1$ start to emerge when $\Omega>\Omega_1$.  
The plot of $\sqrt{\langle M_0^2\rangle}$ versus $\Omega$ in Fig. 4(b) displays a kink at $\Omega=0.20\pm0.02E_{\text{r}}$, fully consistent with the known zero-temperature transition. This provides a benchmark of our finite temperature phase boundary.

Thus, our measurements support the scenario that the phase boundary bends toward the ST phase side, although more complicated scenario can not be completely ruled out. Here we present a simple and quite general symmetry argument to explain why the MG phase is more favorable
than ST phase as temperature increases. Since ST phase breaks both phase symmetry and spatial translational
symmetry along $\hat{x}$, there will be two linear Goldstone modes located at $\pm k_{\text{min}}$ \cite{yunli2013},
as schematically shown in Fig.~5 insets. While MG phase breaks only one continuous phase symmetry,
there is only one linear Goldstone mode. For instance, if atoms condense in $\psi_{\text{R}}$ located at $k_{\text{min}}$,
the spectrum behaves linearly around $k_{\text{min}}$, while remains quadratic around $-k_{\text{min}}$ with
a vanishingly small roton gap in critical regime \cite{Strigari2012}, as schematically shown in insets of Fig.~5.
Thus the MG phase has larger low-energy density-of-state compared to the ST phase, which means the MG phase
can gain more entropy from thermal fluctuations and becomes more favorable.

We then move to the regime with $\Omega_1<\Omega<\Omega_2$ where the ground state is the MG phase.
Here the low-temperature phase exhibits both Bose condensation which breaks $U(1)$ phase symmetry
and spontaneous magnetization which breaks $Z_2$ symmetry.
The question is whether these two different symmetry breaking occur at one single phase transition or
two separated phase transitions. This issue can be addressed in our experiment
since condensate fraction and magnetization can be measured simultaneously.

Here we choose $\Omega=0.6E_{\rm r}$ and $\Omega=3.6E_{\rm r}$.
As shown in Fig.~6(a), we find that when condensate just starts to appear and the condensate fraction $f<0.05$,
the histogram graph clearly shows two peaks located at condensate magnetization $M$ around $\pm 0.95$ for 
$\Omega=0.6E_{\rm r}$ and $M$ around $\pm 0.55$ for $\Omega=3.6E_{\rm r}$.  
Thus, even though we cannot unambiguously conclude that there is only one single phase transition,
at least it shows Bose condensation and magnetization occur in a very narrow temperature window.
Unlike for $\Omega<\Omega_1$, in this regime the double-peak structure remains 
to lower temperature, which is consistent with the MG ground states. 
While at the lowest temperature, the double peak structure is not profound for $\Omega=0.6E_{\text{r}}$ (as shown in Fig. 4(a)). However no signature of phase transition as lowering temperature has been found (for details, see Supplementary material). The suppression of double peak structure may be due to the formation of the ``magnetic domains" in the system, that is, both spin up and spin down atoms could stay in the trap while they are phase separated. 
This is because at very low temperature, it requires longer time to reach global thermal equilibrium than in the higher temperature. Therefore more domains will be formed at the very low temperature than in the higher temperature. This is also consist with the experiment of Lin et.al  \cite{Lin2011} in very low temperature.

However, there is another possibility whether magnetization will occur above condensation temperature, i.e. in thermal gases. 
To examine this possibility, the histograms of magnetization of thermal atoms are shown in Fig. 6(b). 
For each of the coupling strength ($\Omega=0.1E_{\text{r}}, 0.6E_{\text{r}}, 3.6E_{\text{r}}$) and temperature range ($T>T_{\text{c}}, T\approx T_{\text{c}}$), a single narrow peak at $M_{\text{th}}=0$ is shown, which means that the thermal atoms are non-magnetic over large temperature range across $T_{\text c}$ and such a possibility is ruled out. Theoretically, spontaneous magnetization in thermal gas is due to unequal interaction strength between different spin species. For $^{87}\text{Rb}$ atoms, the difference of the s-wave scattering length are tiny and the density of a thermal gas is very low, it is consistent with theoretical expectation that the thermal gas for $T>T_{\text{c}}$ will not display magnetization.

In summary, we experimentally map out the finite-temperature phase diagram of a SO coupled Bose gas of $^{87}\text{Rb}$ atoms realized
by Raman coupling scheme and determine several key features at finite temperature.
These results demonstrate the true power of quantum simulation to guide our understanding, enable us to reveal more interesting critical phenomena for phase transition
between two different types of superfluids.  
Our method can also be applied to similar system with other atoms, such as Dysprosium, where the phase diagram maybe qualitatively different.
Moreover, both the equilibrium and the dynamical behavior of superfluidity in the critical regime are intriguing subjects for future studies. These studies will greatly enrich our knowledge of superfluidity with rich internal structures.

\noindent\textbf{Methods}\\
\noindent\textbf{Preparation and Measurement}.

The setup of this experiment is the same as in our previous work \cite{jinyi2012}.
The $^{87}$Rb atoms are trapped and cooled in an optical dipole trap.
A pair of Raman lasers with the wavelength of $\lambda=803.2$ nm and an incident angle of $105^{\circ}$ in the $x$-$y$ plane
is applied to couple the internal states of $F=1$ manifold.
A bias magnetic field $B=8.4$ Gauss along $z$-direction is applied to generate the Zeeman splitting.
Here the quadratic Zeemen shift $\epsilon=10.14$KHz,
which is $4.6$ times of the recoil energy $E_{\rm r}=2\pi\times2.21\text{KHz}$.
It effectively suppresses $|m_{F}=1\rangle$ state and SO coupling is generated
between $|m_{F}=-1\rangle$ as $|\uparrow\rangle$ and $|m_{F}=0\rangle$ as $|\downarrow\rangle$.

The SO coupled Bose gas at any finite temperature is prepared as following in our experiment.
A single-component Bose gas at $|m_{\text{F}}=-1\rangle$ state is first prepared around $330$ nK with
atom number of $1\sim 2\times 10^{6}$ in an optical dipole trap. Then Raman lasers are adiabatically turned on, normally from $100$ ms to $500$ ms,
which ensures that atoms are loaded into the lower-energy dressed state with SO coupling.
Contrary to the previous experiment where Raman lasers are turned on when condensate is already formed,
in this case, atoms are still thermal.
At the same time, further evaporative cooling is performed to lower the temperature for another $2$ seconds, followed by holding the trap depth
for additional $500$ ms to reach equilibrium. 

For detection, the dipole trap and Raman lasers are switched off suddenly in $1\rm{\mu s}$,
and the atoms are projected back to the bare states ($|m_{\rm{F}}=-1\rangle$ and $|m_{\rm{F}}=0\rangle$ states).
A time-of-flight (TOF) absorption image is taken after 24 ms, with a gradient magnetic field alone $z$-axis to
separate $|m_{\rm{F}}=-1\rangle$ and $|m_{\rm{F}}=0\rangle$ states in vertical direction.
The image resolves the spin and the momentum distribution of the atoms simultaneously.

\noindent\textbf{Heating rate of the dipole trap and the Raman lasers}.

The limitation in the Raman induced spin-orbit coupling is the heating effect. 
This prevent us to reach very low temperature and get high condensate fraction for large $\Omega$. 
The heating rate of the dipole trap is measured to be 18 nK/s, mainly due to the photon scattering and the intensity noise of the dipole trap. 
The heating rate of the Raman lasers with $\delta=0$ is measured to be 180 nK/s for $\Omega=1.0 E_{\text{r}}$, 
which comes from the two photon process with momentum transfer. 
It is about an order of magnitude higher than that induced by photon scattering.

\noindent \textbf{Acknowledgement}

\noindent We acknowledge insightful discussions with Cheng Chin and Tin-Lun Ho. S. C. thanks Bo Zhao for his carefully reading the manuscript.
This work has been supported by the NNSF of China, the CAS, the National Fundamental Research Program
(under Grant No.  2011CB921300, No. 2011CB921500), NSERC and Tsinghua University Initiative Scientific Research Program.

\noindent \textbf{Authors contribution}

\noindent S. C. and J. -W. P. plan and supervise the project. S. -C. J., J. -Y. Z, Z. -D. D and S. C. perform the experiment, L. Z., W. Z., Y. -J. D., and H. Z. give the theoretical support, all the authors contributed to analyze the data and write the manuscript.

\noindent \textbf{Additional information}

\noindent The authors declare no competing financial interests. \\
\noindent $*$ These authors contribute equally in this work. \\
\noindent $\dagger$ Correspondence and requests for materials should be addressed to S. Chen(shuai@ustc.edu.cn) or J.-W. Pan(pan@ustc.edu.cn).

\clearpage

\bibliographystyle{naturemag}
\bibliography{Reference}

\noindent\textbf{Reference}
\begin{thebibliography}{99}

\bibitem{topological_insulator} Hasan, M. Z. \& Kane, C. L. Topological insulators. {\em Rev. Mod. Phys.} {\bf 82}, 3045--3067 (2010).

\bibitem{topological_insulator2} Qi, X.-L. \& Zhang, S.-C. Topological insulators and superconductors. {\em Rev. Mod. Phys.} {\bf 83}, 1057--1110 (2011).

\bibitem{Lin2011}  Lin, Y.-J., Jim\'enez-Garc\'ia, K. \& Spielman, I. B. Spin-orbit-coupled Bose-Einstein condensates. {\em Nature} {\bf 471}, 83--86 (2011).

\bibitem{JZhang2012} Wang, P. {\em et al.} Spin-orbit coupled degenerate Fermi gases. {\em Phys. Rev. Lett.} {\bf 109}, 095301 (2012).

\bibitem{MZwielein2012} Cheuk, L. M. {\em et al.} Spin-injection spectroscopy of a spin-orbit coupled Fermi gas. {\em Phys. Rev. Lett.} {\bf 109}, 095302 (2012).

\bibitem{jinyi2012} Zhang, J.-Y. {\em et al.} Collective dipole oscillation of a spin-orbit coupled Bose-Einstein condensate. {\em Phys. Rev. Lett.} {\bf 109}, 115301 (2012).

\bibitem{Spielman2012} Williams, R. A. {\em et al.} Synthetic partial waves in ultracold atomic collisions. {\em Science} {\bf 335}, 314--317 (2012).

\bibitem{chuanwei2013} Qu, C., Hamner, C., Gong, M., Zhang, C. \& Engels, P. Observation of Zitterbewegung in a spin-orbit-coupled Bose-Einstein condensate. {\em Phys. Rev. A} {\bf 88}, 021604(R) (2013).


\bibitem{long2013} Zhang, L. {\em et al.} Stability of excited dressed states with spin-orbit coupling. {\em Phys. Rev. A} {\bf 87}, 011601(R) (2013).

\bibitem{huizhai2010} Wang, C., Gao, C., Jian, C.-M. \& Zhai, H. Spin-orbit coupled spinor Bose-Einstein condensates. {\em Phys. Rev. Lett.} {\bf 105}, 160403 (2010).

\bibitem{Ho2011} Ho, T.-L. \& Zhang, S. Bose-Einstein condensates with spin-orbit interaction. {\em Phys. Rev. Lett.} {\bf 107}, 150403 (2011).

\bibitem{yunli2012} Li, Y., Pitaevskii, L. P. \& Stringari, S. Quantum tricriticality and phase transitions in spin-orbit coupled Bose-Einstein condensates. {\em Phys. Rev. Lett.}, {\bf 108}, 225301 (2012).

\bibitem{Congjun} Wu, C.-J., Mondragon-Shem, I. \& Zhou, X.-F. Unconventional Bose-Einstein condensations from spin-orbit coupling. {\em Chin. Phys. Lett.} {\bf28}, 097102 (2011).

\bibitem{Jian} Jian, C.-M. \& Zhai, H. Paired superfluidity and fractionalized vortices in systems of spin-orbit coupled bosons. {\em Phys. Rev. B} {\bf 84}, 060508(R) (2011).

\bibitem{Lamacraft} Gopalakrishnan, S., Lamacraft, A. \& Goldbart, P. M. Universal phase structure of dilute Bose gases with Rashba spin-orbit coupling. {\em Phys. Rev. A} {\bf 84}, 061604(R) (2011).

\bibitem{Baym} Ozawa T. \& Baym, G. Stability of Ultracold Atomic Bose Condensates with Rashba Spin-Orbit Coupling against Quantum and Thermal Fluctuations. {\em Phys. Rev. Lett.} {\bf109}, 025301 (2012).

\bibitem{Cui} Cui, X. \& Zhou, Q. Enhancement of condensate depletion due to spin-orbit coupling. {\em Phys. Rev. A} {\bf87}, 031604(R) (2013).


\bibitem{Zheng} Zheng, W., Yu, Z.-Q., Cui, X. \& Zhai, H. Properties of Bose gases with Raman-induced spin-orbit coupling.
{\em J. Phys. B}, {\bf 46}, 134007 (2013).

\bibitem{Kamenev} Sedrakyan, T. A., Kamenev, A. \& Glazman, L. I. Composite fermion state of spin-orbit coupled Bosons {\em Phys. Rev. A} {\bf 86}, 063639 (2012).

\bibitem{Biao} Zhu, Q., Zhang, C. \& Wu, B. Exotic superfluidity in spin-orbit coupled Bose-Einstein condensates. {\em Europhys. Lett.} {\bf 100}, 50003 (2012).

\bibitem{Strigari2012} Martone, G. I., Li, Y., Pitaevskii, L. P. \& Stringari, S. Anisotropic dynamics of a spin-orbit coupled Bose-Einstein condensate. {\em Phys. Rev. A} {\bf 86}, 063621 (2012).


\bibitem{Z2} Mukerjee, S., Xu, C. \& Moore, J. E. Topological defects and the superfluid transition of the $s=1$ spinor condensate in two dimensions. {\em Phys. Rev. Lett.} {\bf 97}, 120406 (2006).

\bibitem{Lamacraft1} James, A. J. A. \& Lamacraft, A. Phase diagram of two-dimensional polar condensates in a magnetic field. {\em Phys. Rev. Lett.} {\bf 106}, 140402 (2011).

\bibitem{Lamacraft2} Shi, Y., Lamacraft, A. \& Fendley, P. Boson pairing and unusual criticality in a generalized XY model. {\em Phys. Rev. Lett.} {\bf 107}, 240601 (2011).

\bibitem{spielmanPRA2009} Spielman, I. B. Raman process and effective gauge potentials. {\em Phys. Rev. A} {\bf 79}, 063613 (2009).

\bibitem{He-3} Wheatley, J. C. Experimental properties of superfluid $^3$He. {\em Rev. Mod. Phys.} {\bf 47}, 415--470 (1975).

\bibitem{mag} de la Cruz, C. {\em et al.} Magnetic order close to superconductivity in the iron-based layered
LaO$_{1-\text{x}}$F$_\text{x}$FeAs system. {\em Nature} {\bf 453}, 899--902 (2008).

\bibitem{mag1} Fang, C., Yao, H., Tsai, W.-F., Hu, J.-P., \& Kivelson, S. A. Theory of electron nematic order in LaFeAsO.
{\em Phys. Rev. B} {\bf 77}, 224509 (2008).

\bibitem{mag2} Xu, C.-K., M\"uller, M., \& Sachdev, S. Ising and spin orders in the iron-based superconductors.
{\em Phys. Rev. B} {\bf 78}, 020501(R) (2008).




\bibitem{yunli2013} Li, Y., Martone, G. I., Pitaevskii, L. P. \& Stringari, S. Superstripes and the excitation spectrum of a spin-orbit-coupled Bose-Einstein condensate. 
{\em Phys. Rev. Lett.} {\bf 110}, 235302 (2013). 



\end{thebibliography}

\clearpage

\textbf{Figure 1: Zero-temperature phase diagram and three scenarios
        for finite-temperature phase diagram for SO coupled Bose gas.}
\textbf{a.} Single-particle dispersion and zero-temperature phase diagram as a function of Raman coupling $\Omega$,
            which shows stripe phase, plane wave phase and non-magnetic phase as $\Omega$ increases.
\textbf{b1-b3.} Three different scenarios of finite-temperature phase diagram in terms of $\Omega$ and temperature $T$.

\textbf{Figure 2: Critical temperature $T_\text{c}$ of spin-orbit coupled Bose gas.}
\textbf{a}. Spin-resolved time-of-flight image of momentum distribution. At $T=63~{\text{nK}}$, the population imbalance of the condensate shows that it is in magnetized phase.
\textbf{b}. Condensate fraction $f$ as a function of temperature $T$. 
\textbf{c}. Ratio $T_{\text{c}}/T_{\text{c}}^{({\rm s})} $ of the measured critical temperature of SO coupled Bose gas
            to that of single-component Bose gas with same density. The experimental data are shown by black squares
            with error bar. Two dashed lines correspond to $1$ and $(1/2)^{1/3}$, respectively.
\textbf{d}. Interaction shift of critical temperature $\Delta T_{\text{c}}/T_{\text{c}}$,
            where $\Delta T_{\text{c}}=T_{\text{c}}-T^{(0)}_{\text{c}}$, and $T^{(0)}_{\text{c}}$ is
            non-interacting critical temperature calculated for SO coupled Bose gas for same density and trap frequency.
            The red and blue solid lines are the theoretical curves with atom number $N=1\times 10^5$,
             trapping frequency $\bar{\omega}=30$ Hz and $N=1\times 10^6, \bar{\omega}=70$ Hz, respectively.
 	   All the error bars are the standard errors which transferred from the measurement of atom number $N$, $N_0$ and temperature $T$.
             
\textbf{Figure 3: Histogram of condensate magnetization for $\Omega<\Omega_1$.}
\textbf{(a-c)}. Evolution of magnetization histogram as temperature is lowered
                (i.e. increasing of condensate fraction $f$) with $\Omega=0.10E_{\rm r}$ (a1-a6),
                $\Omega=0.15E_{\rm r}$(b1-b6) and $\Omega=0.18E_{\rm r}$(c1-c6).
                Dashed line indicates the ``boundary" where the "double-peak" at $M=\pm 1$ disappears.
\textbf{(d)}.  $\sqrt{\langle M_0^2\rangle}$ as a function of $f$ for $\Omega=0.10E_{\rm r}$ (d),
                $\Omega=0.15E_{\rm r}$ (e) and $\Omega=0.18E_{\rm r}$ (f), respectively. The error bars are statistic errors. 

\textbf{Figure 4: Phase transition between ST condensate and MG condensate at very low temperature}
\textbf{(a)}. Magnetization histograms with varying $\Omega$ for nearly pure condensate ($T<20 \text{nK}$). 
\textbf{(b)}.  $\sqrt{\langle M_0^2\rangle}$ as a function of $\Omega$ at $T<20 \text{nK}$, the error bars are the standard statistics error transferred from the measurement. The fitting curve is applied to guide the eye.               

\textbf{Figure 5: Finite-temperature phase diagram of spin-orbit coupled Bose gas}. Finite-temperature phase diagram of spin-orbit coupled bosons, the error bars are transferred from the measured $f-T$ relation as Fig. 2(b). 
insets: Schematic low-energy spectrum for stripe and plane wave phase.

\textbf{Figure 6: Formation of magnetic order for $\Omega_1<\Omega<\Omega_2$}.
\textbf{(a)}. Evolution of magnetization histogram as temperature is lowered (i.e. increasing of condensate fraction $f$)
with $\Omega=0.6E_{\rm r}$, $\Omega=3.6E_{\rm r}$.
\textbf{(b)}. Evolution of magnetization histograms of thermal atoms with temperature range cover across $T_{\text{c}}$ at $\Omega=0.1E_{\text{r}}$,
		$\Omega=0.6E_{\text{r}}$ and $\Omega=3.6E_{r}$.

\clearpage

\begin{figure}[t!]\label{phasediagram}
\begin{center}
\includegraphics[width=1.0\linewidth]{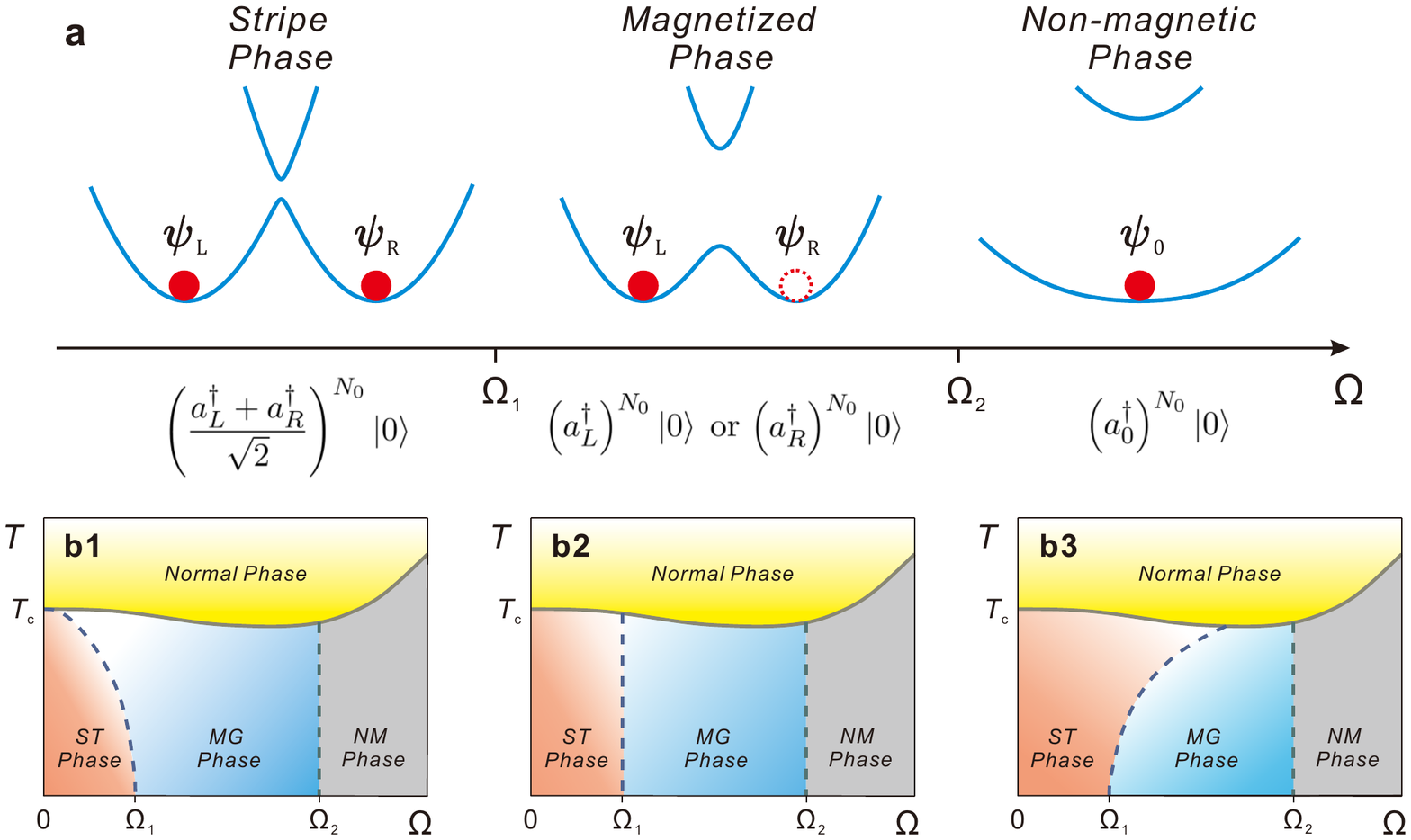}
\vspace{-0.5cm}
\caption{\textbf{Zero-temperature phase diagram and three scenarios for
finite-temperature phase diagram for spin-orbit coupled Bose gas }  %
\label{fig_phasediagram}}
\end{center}
\end{figure}

\begin{figure}[t!]
\begin{center}
\includegraphics[width=1.0\linewidth]{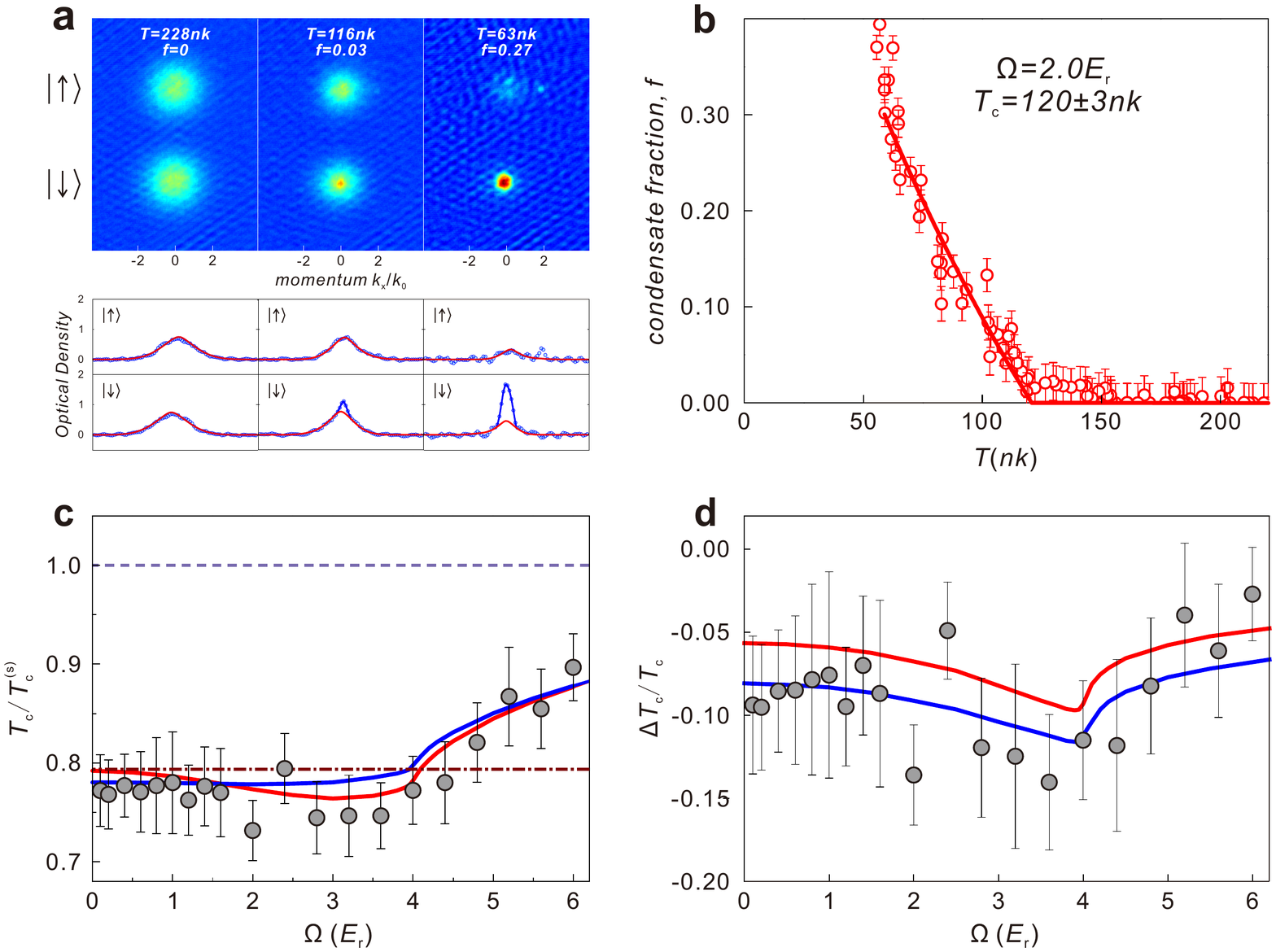}
\vspace{-0.5cm}
\caption{\textbf{Critical temperature $T_\text{c}$ of spin-orbit coupled Bose gas}  %
\label{Fig_$T_{c}$}}
\end{center}
\end{figure}

\begin{figure}[t!]
\begin{center}
\includegraphics[width=1.0\linewidth]{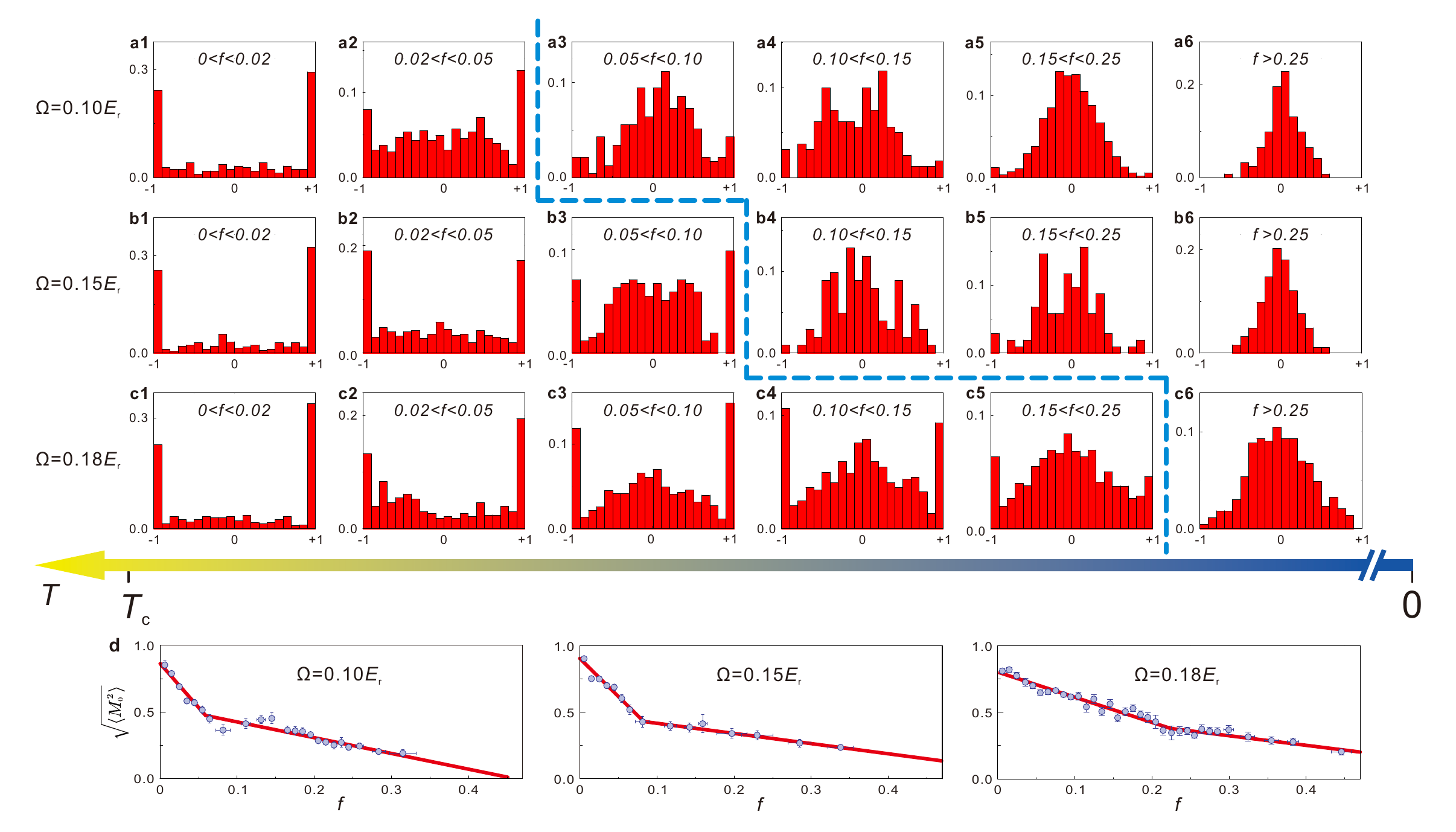}
\vspace{-1.0cm}
\caption{\textbf{Histogram of condensate magnetization for $\Omega<\Omega_1$.}  %
\label{Fig_Theory}}
\end{center}
\end{figure}

\begin{figure}[t!]
\begin{center}
\includegraphics[width=1.0\linewidth]{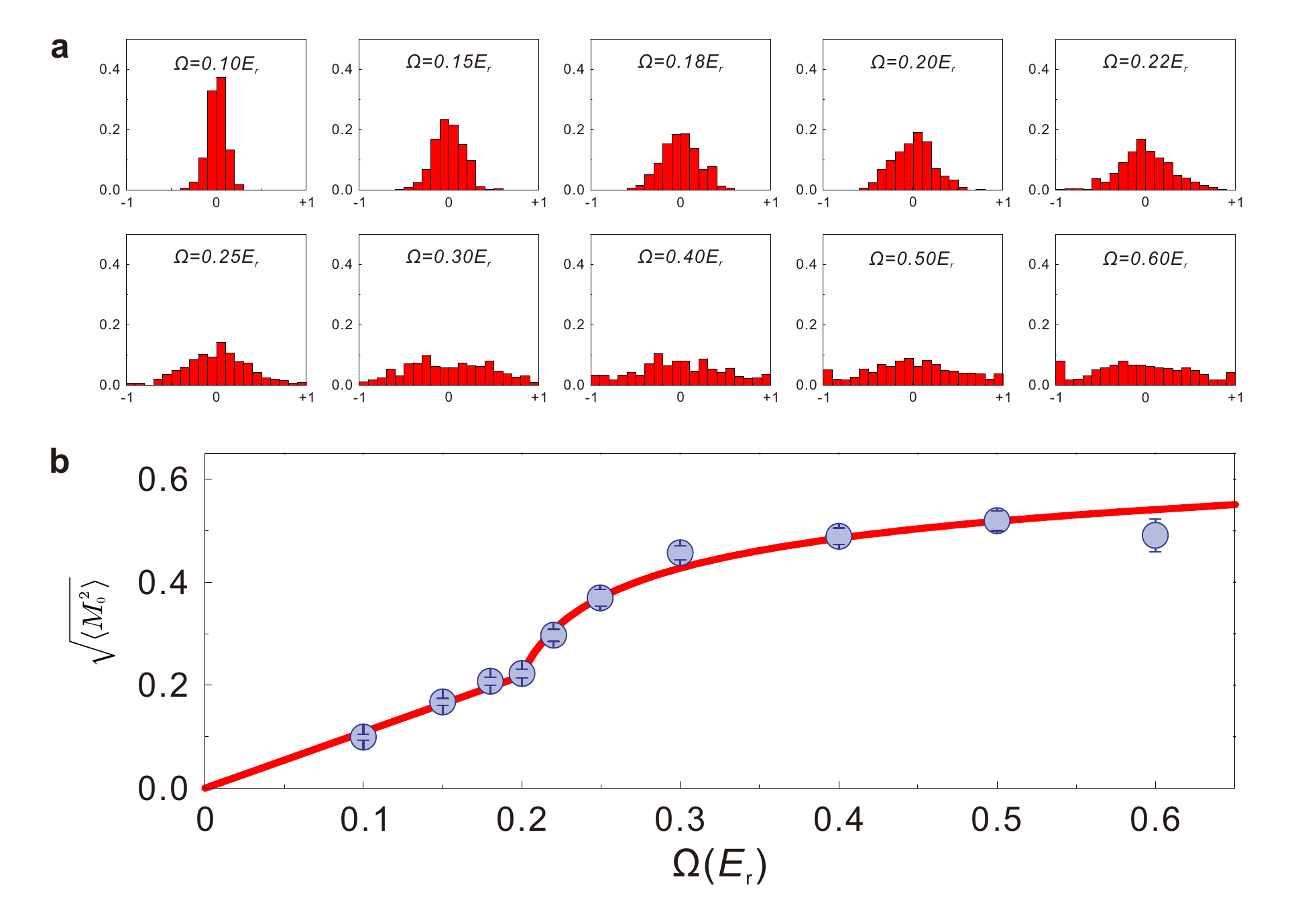}
\vspace{-0.5cm}
\caption{\textbf{Phase transition between ST condensate and MG condensate at very low temperature}  %
\label{Fig_Data}}
\end{center}
\end{figure}

\begin{figure}[t!]
\begin{center}
\includegraphics[width=1.0\linewidth]{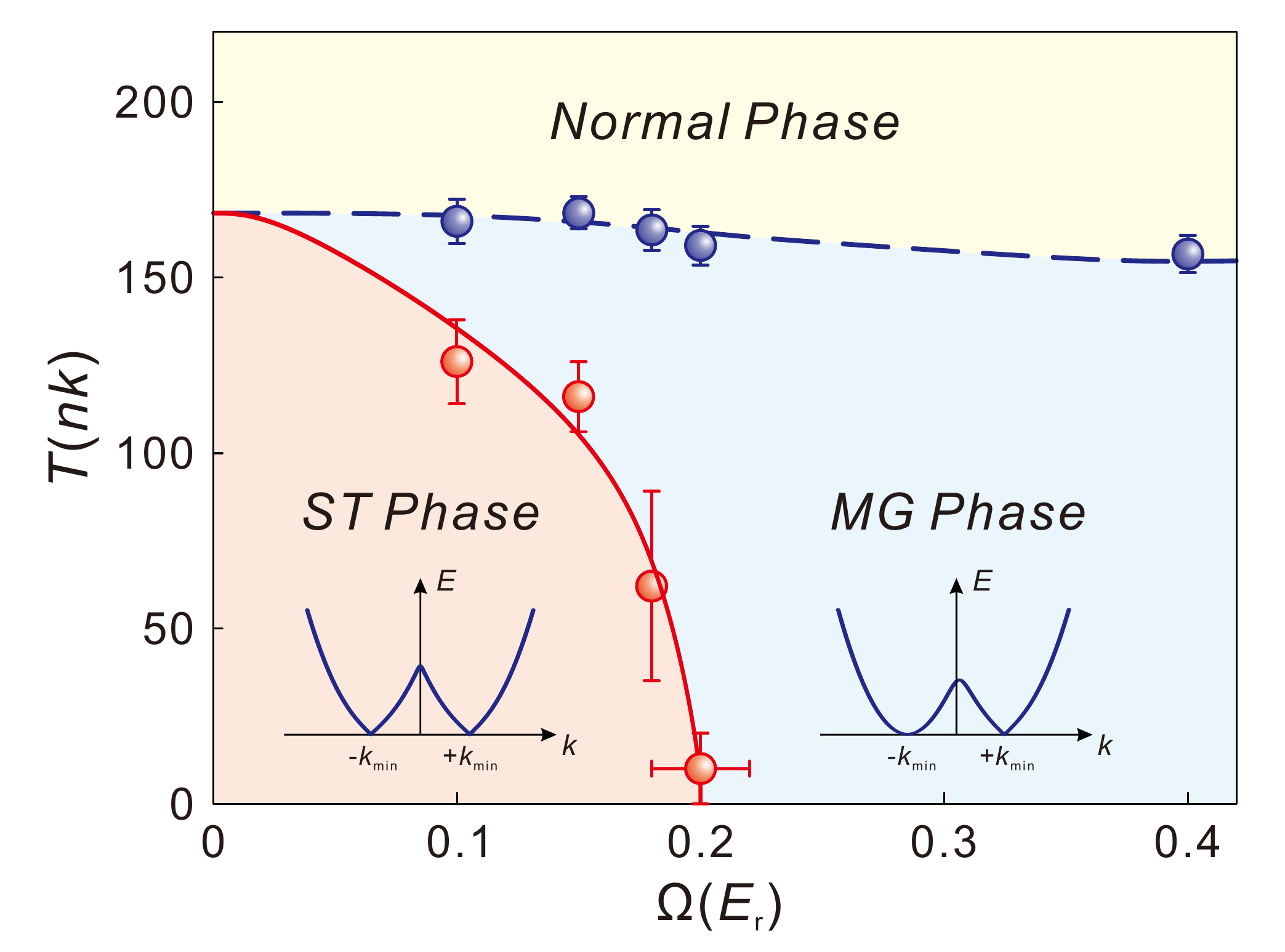}
\vspace{-0.5cm}
\caption{\textbf{Finite-temperature phase diagram of spin-orbit coupled Bose gas}   %
\label{Fig_Data}}
\end{center}
\end{figure}

\begin{figure}[t!]
\begin{center}
\includegraphics[width=1.0\linewidth]{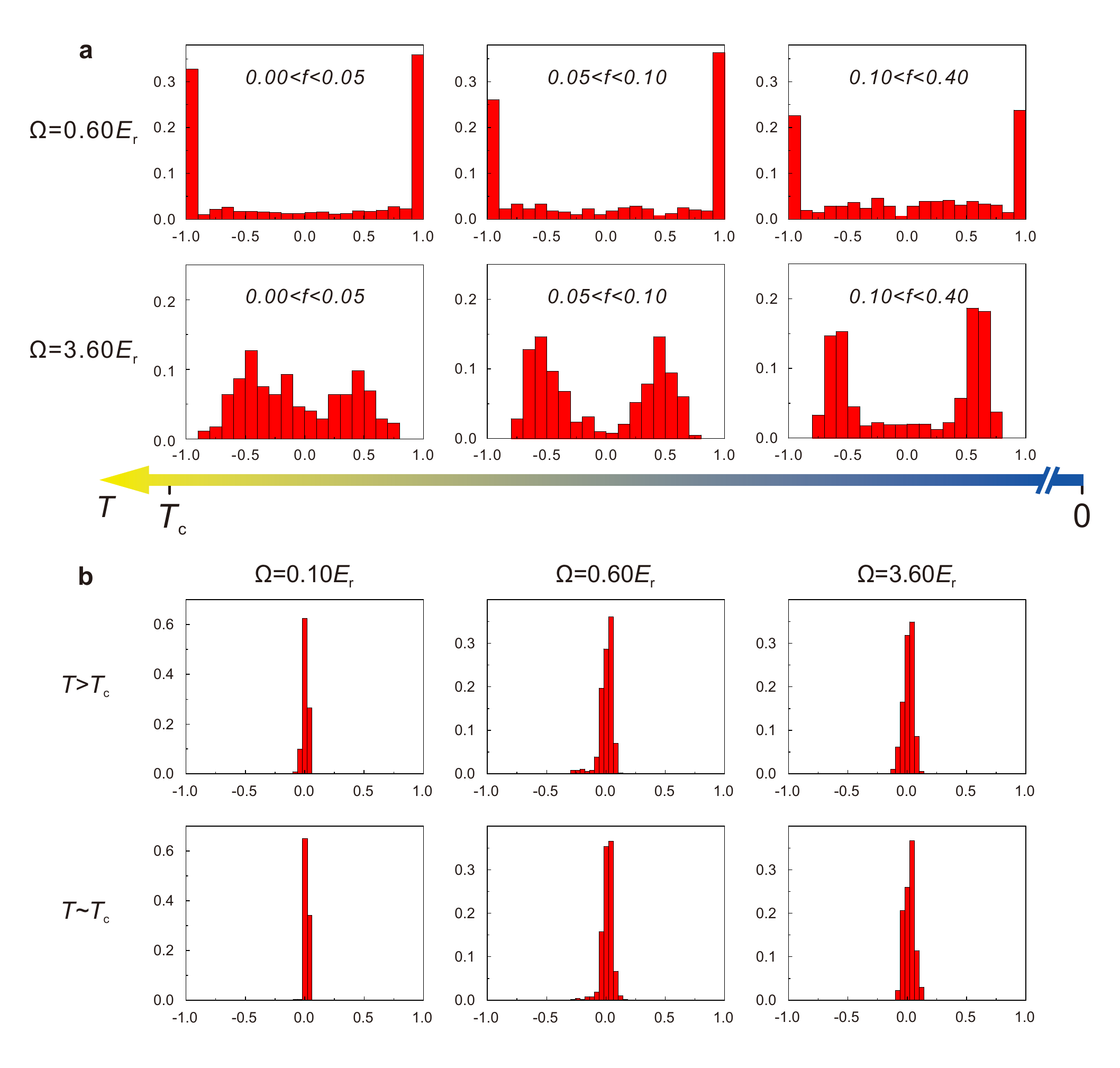}
\vspace{-0.5cm}
\caption{\textbf{Formation of magnetic order for $\Omega_1<\Omega<\Omega_2$}  %
\label{Fig_$T_{c}$}}
\end{center}
\end{figure}

\clearpage





\end{document}